\documentclass[showpacs,reprint,twocolumn,floatfix]{revtex4-1}

\usepackage{graphicx,float,amsmath}
\usepackage{natbib}

\begin{document}

\title{Geometric and chemical components of the giant piezoresistance in silicon nanowires}

\author{M.M. McClarty$^1$}
\author{N. Jegenyes$^1$}
\author{M. Gaudet$^2$}
\thanks{Current address: Fraunhofer Institute for Photonic Microsystems, Dresden, Germany}
\author{C. Toccafondi$^3$}
\author{R. Ossikovski$^3$}
\author{F. Vaurette$^2$}
\author{S. Arscott$^2$}
\email{steve.arscott@iemn.univ-lille1.fr}
\author{A.C.H. Rowe$^1$}
\email{alistair.rowe@polytechnique.edu}

\affiliation{$^1$Physique de la Mati\`ere Condens\'ee, Ecole Polytechnique, CNRS, 91128 Palaiseau, France}

\affiliation{$^2$Institut d'Electronique, de Micro\'electronique et de Nanotechnologie (IEMN), Universit\'e de Lille, CNRS, Avenue Poincar\'e, Cit\'e Scientifique, 59652 Villeneuve d'Ascq, France}

\affiliation{$^3$LPICM, Ecole Polytechnique, CNRS, 91128 Palaiseau, France}

\begin{abstract}
A wide variety of apparently contradictory piezoresistance (PZR) behaviors have been reported in p-type silicon nanowires (SiNW), from the usual positive bulk effect to anomalous (negative) PZR and giant PZR. The origin of such a range of diverse phenomena is unclear, and consequently so too is the importance of a number of parameters including SiNW type (top down or bottom up), stress concentration, electrostatic field effects, or surface chemistry. Here we observe all these PZR behaviors in a single set of nominally p-type, $\langle 110 \rangle$ oriented, top-down SiNWs at uniaxial tensile stresses up to 0.5 MPa. Longitudinal $\pi$-coefficients varying from $-800\times10^{-11}$ Pa$^{-1}$ to $3000\times10^{-11}$ Pa$^{-1}$ are measured. Micro-Raman spectroscopy on chemically treated nanowires reveals that stress concentration is the principal source of giant PZR. The sign and an excess PZR similar in magnitude to the bulk effect are related to the chemical treatment of the SiNW.

\end{abstract}
\maketitle

Mechanical stress modifies the electronic structure of solids and can lead to a change in their resistivity \cite{rowe2014}. This effect, called piezoresistance (PZR), is well known in bulk crystalline silicon \cite{smith1954} and is widely exploited in order to improve the speed, gain and symmetry of modern CMOS microelectronic devices \cite{thompson2006}, as well as enabling silicon based MEMS sensor technologies \cite{du2007}. Silicon nanowires (SiNW) have attracted attention for their PZR because of reports of giant or anomalous effects \cite{he2006, lugstein2010, neuzil2010, kang2012, kang2012b} that differ either in sign or magnitude from the bulk PZR. The physical mechanism responsible for these observations is unclear. While the lateral dimensions of tested SiNWs are typically too large for quantum confinement to play a role \cite{dorda1971}, it was noted that giant PZR is associated with partial depletion of free charge carriers \cite{rowe2008}, achieved either by reducing the doping density \cite{he2006} or by gating into the sub-threshold region \cite{neuzil2010,kang2012,kang2012b}. This suggests an electrostatic origin for the giant PZR, for example due to electromechanically active interface states \cite{rowe2008}. Although there is some evidence for such a component in the anomalous PZR at very high stresses \cite{jang2014,winkler2015}, in most cases PZR close to the bulk value is observed \cite{toriyama2003, reck2008, milne2010, barwicz2010, koumela2011, bhaskar2013}, even in depleted \cite{milne2010} or gated \cite{bhaskar2013} SiNWs. It is important to clarify this situation in part because a novel surface electromechanical phenomenon may be involved, but also in the context of sensing and strain effects on the electronic properties of future nanoscale silicon devices.

Here we set out to experimentally determine the role of geometric stress concentration \cite{peterson1975}, particularly in released SiNWs where the largest PZR is claimed \cite{he2006}. The PZR is expressed as the $\pi$-coefficient, \begin{equation} \label{pi} \pi \approx \frac{1}{X_{ext}}\frac{\Delta R}{R_0}, \end{equation} where $R_0$ is the zero stress resistance, $\Delta R$ is the stress-induced resistance change, and $X_{ext}$ is the externally applied stress. In bulk crystalline silicon $\pi$ depends on the crystal direction and doping type \cite{smith1954}. In the $\langle 110 \rangle$ direction of interest here $\pi = \pi_{bulk} = +71 \times 10^{-11}$ Pa$^{-1}$ in p-Si and $\pi = \pi_{bulk} = -31 \times 10^{-11}$ Pa$^{-1}$ in n-Si \cite{smith1954}. In a mechanical constriction such as an individual SiNW, the local stress may be significantly larger than the externally applied value. When stretched along its axis, this can be quantified by a stress concentration factor ($b$) where \begin{equation} \label{b} X = b\times X_{ext} \end{equation} and $X$ is the true stress in the SiNW. According to Eqns. \ref{pi} and \ref{b}, the apparent PZR in a SiNW due solely to stress concentration will then be \begin{equation} \label{effectivepi} \pi_{NW} = b \times \pi_{bulk}, \end{equation} which can be significantly larger than the bulk $\pi$-coefficients if the geometry is such that $b \gg 1$. Finite element calculations \cite{suppinfoapl2016} reveal that although $b$ is largest for small aspect ratio nanowires, it can exceed 10 in suspended, larger aspect ratio nanowires like those studied elsewhere \cite{he2006}.

\begin{figure}[htbp]
\includegraphics[clip,width=8.5 cm] {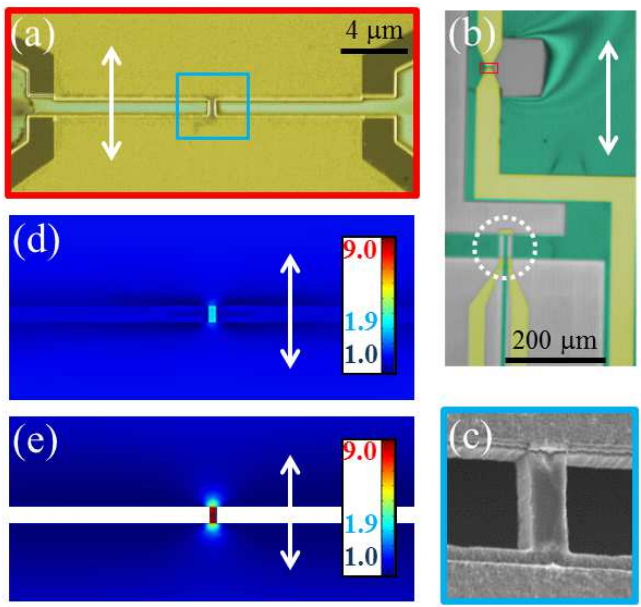}
\caption{Sample details and expected geometric stress concentration factors, $b$. In all images the uniaxial tensile stress is applied parallel to the double headed white arrow, with the colors in the SEM images corresponding to the active layer (gray), the BOX (green), and the Ohmic contacts (yellow). (a) Top view SiNW corresponding to the red box shown in (b). (b) A zoomed out image showing a SiNW (inside red rectangle) and its neighboring strain gage (white dotted circle). (c) A magnified, gray scale image of a SiNW corresponding to the blue box in (a). Finite element calculations showing (d) $b = 1.9$ in a non-released SiNW, and (e) $b = 9$ in a released SiNW.}
\label{sample}
\end{figure}

With this in mind nominally identical top-down SiNWs were fabricated from a silicon-on-insulator (SOI) wafer whose active layer is $d =$ 300 nm thick ($p=2.6\times 10^{13}$ cm$^{-3}$) and whose buried oxide (BOX) is $h = 3$ $\mu$m thick. The active layer was implanted with Boron ($p=3.8\times 10^{18}$ cm$^{-3}$) and SiNWs 300 nm wide and 1 $\mu$m long, oriented parallel to the $\langle 110 \rangle$ crystal direction, were defined in the active layer using electron beam lithography and chlorine based dry etching (see Fig. \ref{sample}(a)). Each SiNW is ohmically contacted with lithographically defined PtSi contacts (shown in yellow in Fig. \ref{sample}). In this configuration finite element modelling of the stress/strain relationship reveals that, when strained parallel to the double-headed white arrow in Fig. \ref{sample}, $b$ = 1.9 (see Fig. \ref{sample}(b)). $X_{ext}$ can be measured \textit{in situ} by simultaneously monitoring the PZR of a macroscopically large, p-type strain gage defined in the active layer close to the SiNW (see Fig. \ref{sample}(c)). If the apparent SiNW PZR is just the bulk PZR boosted by stress concentration, then $\pi_{NW}/\pi_{bulk}$  should yield $b$ according to Eq. \ref{effectivepi}. When the SiNW is released by etching off the underlying BOX using a concentrated (50 \%) HF etch for 2 minutes 20 seconds (see Fig. \ref{sample}(d)), the local strain is no longer restricted by contact with the BOX and the stress concentration is more effective with $b$ = 9 (see Fig. \ref{sample}(e)). While larger values of $b$ are possible for thinner SiNWs \cite{suppinfoapl2016}, this can be to the detriment of process yield. In total 20 non-released SiNWs were measured for the study, of which 9 were released.

\begin{figure}[htbp]
\includegraphics[clip,width=8.5 cm] {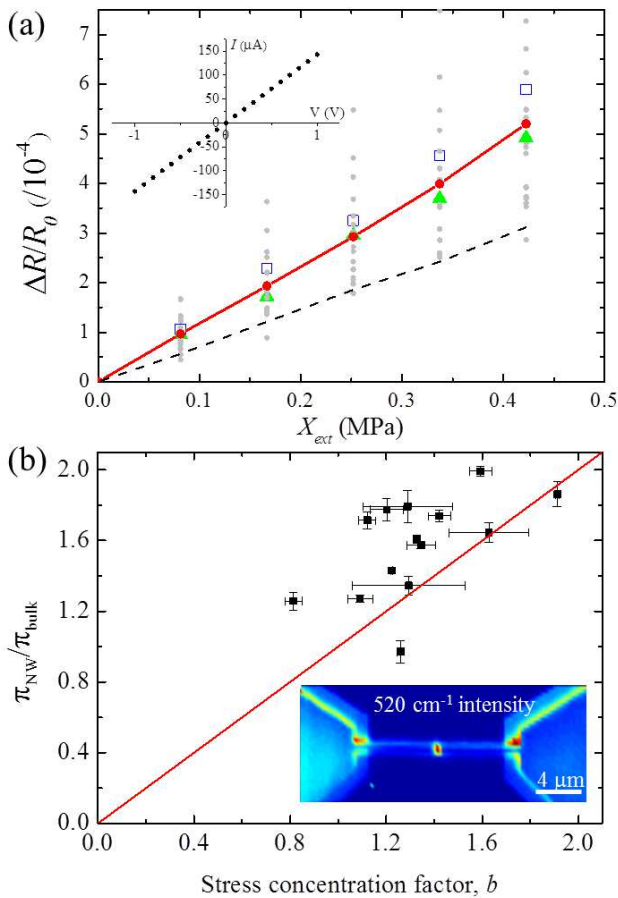}
\caption{(a) PZR of 20 non-released SiNWs (gray points). The linear response is highlighted by two SiNWs whose properties are shown (below) after release (blue squares, green triangles). The average response (shown in red) has a slope $\approx 1.7$ times the gage response (black, dotted line) which is close expectd value, $b =$ 1.9. The inset shows a typical linear IV curve consistent with the nominal doping density. (b) The ratio $\pi_{NW}/\pi_{bulk}$ for each SiNW plotted against $b$ as measured using micro-Raman spectroscopy (see inset). The very good correlation (c.f. red line of slope 1) confirms that the apparent PZR is due to stress concentration.}
\label{unreleased}
\end{figure}

Figure \ref{unreleased}(a) shows the $X_{ext}$ dependence of $\Delta R/R_0$ obtained on the non-released SiNWs. Results from individual SiNWs are shown as gray dots, with two individual cases used to highlight the typical linear response (green triangles and blue squares). The red line corresponds to the average response over the 20 measured SiNWs while the typical response from the strain gages is shown by the dashed, black line. The slope of the red line, proportional to $\pi_{NW}$, is approximately 1.7 times that of the dashed, black line whose slope is proportional to $\pi_{bulk}$, which is close to the expected value, $b =$ 1.9 (Fig. \ref{sample}(b)). Thus the apparent PZR, $\pi_{NW} \approx +120 \times 10^{-11}$ Pa$^{-1}$, is due to stress concentration. A similar result was obtained in the first work on PZR in SiNWs \cite{toriyama2003}.

The actual value of $b$ for each SiNW is determined by their individual, imperfect geometries and can be evaluated using micro-Raman imaging of the shift in the 520 cm$^{-1}$ silicon peak with an applied mechanical stress \cite{ossikovski2008, suppinfoapl2016}. Despite the size of the SiNWs, whose lateral dimension is close to the diffraction limit of the $\times$100 microscope objective used with a 532 nm excitation, it is possible to clearly identify the SiNW in an image of the 520 cm$^{-1}$ peak intensity (see inset, Fig. \ref{unreleased}(b)). The horizontal axis of Fig. \ref{unreleased}(b) shows this ratio, which varies from about 1.2 to 2, again close to the expected value of 1.9. When the ratio $\pi_{NW}/\pi_{bulk}$ for each SiNW is plotted against $b$, Fig. \ref{unreleased}(b) is obtained. There is a very good correlation between the PZR ratio and $b$ (see comparison with the red line of slope 1), confirming the validity of Eq. \ref{effectivepi} and the geometric nature of the apparent PZR. This suggests that when released these same SiNWs should show the predicted 9-fold increase in the PZR.

\begin{figure}[htbp]
\includegraphics[clip,width=8.5 cm] {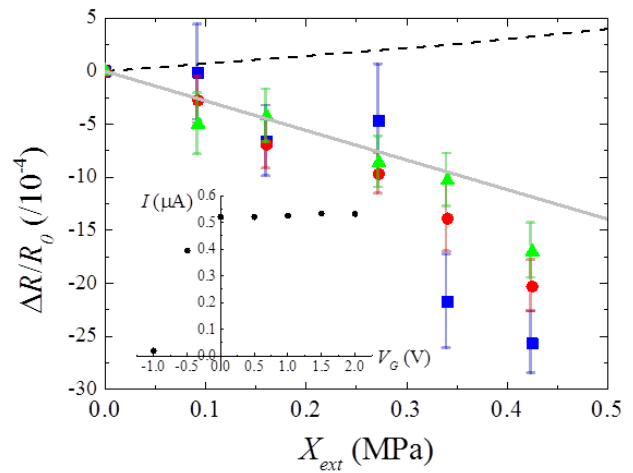}
\caption{Anomalous (negative) PZR of naturally re-oxidized, released SiNWs whose magnitude is significantly larger than in bulk p-type silicon (dashed, black line). The data (green triangles) comes from the SiNW whose pre-release response is shown with the same symbol in Fig. \ref{unreleased}(a). Gating experiments (inset) reveal that the SiNWs are now n-type, possibly due to acceptor neutralization by HF. This is consistent with the sign of the PZR. The known bulk n-type PZR multiplied by $b$ = 9 yields  of $\pi_{NW} = -279\times 10^{-11}$ Pa$^{-1}$ (gray line) which is close to the measured PZR.}
\label{nreleased}
\end{figure}

After the SiNWs are released from the BOX using concentrated HF and then allowed to naturally re-oxidize in air, an unexpected result is obtained. As shown in Fig. \ref{nreleased} the PZR is now negative, consistent with n-type silicon. However, the maximum $\pi_{NW}\approx -800\times10^{-11}$ Pa$^{-1}$ is signficantly larger than the PZR of either n- or p-type (dashed, black line) silicon \cite{smith1954}. This striking change in the PZR is well illustrated by the data (green triangles) taken from a SiNW whose pre-release PZR is shown with the same symbol in Fig. \ref{unreleased}(a). Using  the substrate (handle) of the SOI wafer as a gate, an increasingly negative gate bias reduces the conductivity (see Fig. \ref{nreleased}, inset) which confirms the n-type nature of the SiNWs. Note that although the sign is the same as the anomalous PZR reported elsewhere \cite{lugstein2010, jang2014,winkler2015}, here the stress is significantly lower and there is no indication that these are the same phenomenon. 

The giant magnitude of the PZR can be understood by observing the close agreement between the data and the estimated value for n-Si after taking stress concentration into account, $\pi_{NW} = -9 \times 31 \times 10^{-11} = -279 \times 10^{-11}$ Pa$^{-1}$ (gray line in Fig. \ref{nreleased}). Although Raman measurements on these samples were not possible, this comparison suggests that the true PZR (i.e. that normalized to the local stress in the SiNW) is simply that of bulk n-Si. The more complex question is why the SiNWs, previously p-type with a resistance of a few k$\Omega$ (see inset, Fig. \ref{unreleased}(a)), have switched type and become more resistive by up to 5 orders of magnitude? A possible explanation is the neutralization of the Boron acceptors by atomic hydrogen which can occur within a few micrometers of the surface \cite{pankove1983, huang1992}, sufficient in the case of SiNWs to neutralize all acceptors. Natural re-oxidation of an HF treated surface is also known to create an interface defects that are principally donor-like \cite{angermann2014}, possibly accounting for the weakly n-type nature of the SiNWs.

\begin{figure}[htbp]
\includegraphics[clip,width=8.5 cm] {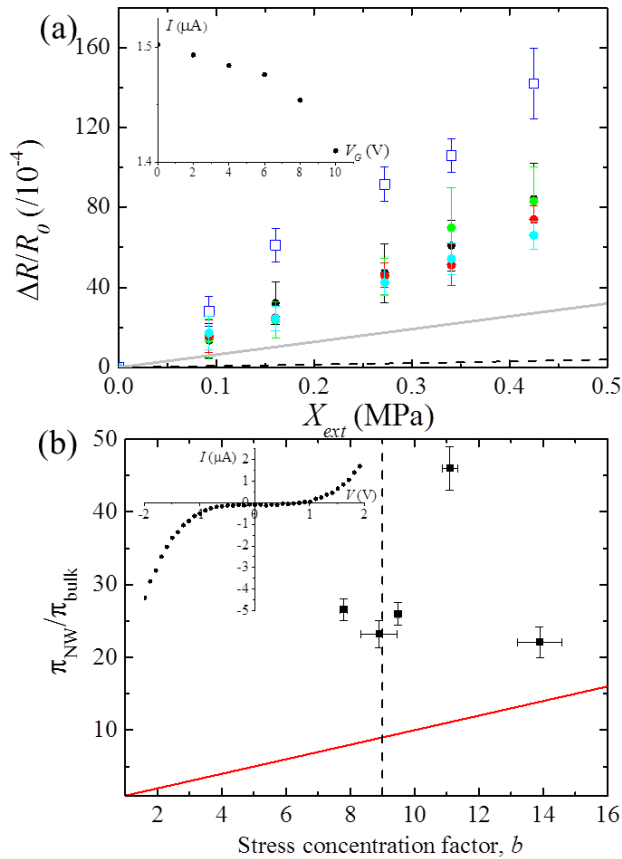}
\caption{PZR of released SiNWs after a 30 second etch in 70 \% HNO$_3$. (a) PZR is positive as expected for p-type silicon but significantly larger (at most $\pi_{NW} \approx +3000\times 10^{-11}$ Pa$^{-1}$) than both $\pi_{bulk}$ (dashed, black line) and $b\times\pi_{bulk}$ with $b = 9$ (solid, gray line). The data (blue squares) comes from the SiNW whose pre-release response is shown with the same symbol in Fig. \ref{unreleased}(a). Gating experiments (inset) confirm the p-type nature of the SiNWs. (b) $\pi_{NW}/\pi_{bulk}$ plotted against $b$ as measured using micro-Raman spectroscopy confirming that $b$ is close to the expected value (vertical, dashed, black line) and that the PZR is above $b\times\pi_{bulk}$ (solid, red line). The IV characteristic (inset) is non-linear indicating partial depletion of the SiNWs.}
\label{preleased}
\end{figure}

A second group of SiNWs was released in the same way. In contrast to the previous set, these SiNWs were then rapidly re-oxidized in 70 \% HNO$_3$ for 30 seconds \cite{he2006} rendering them p-type as shown by the positive PZR and by gating experiments (see Fig. \ref{preleased}(a)). Again, the striking effect of the release and the HNO$_3$ treatment is well illustrated by the data marked with blue squares taken from a SiNW whose pre-release PZR is shown with the same symbol in Fig. \ref{unreleased}(a). The p-type nature of the SiNWs is not due to the Boron acceptors which require a vacuum anneal for re-activation \cite{huang1992}, but is consistent with the creation of acceptor-like interface defects by HNO$_3$ oxidation \cite{yuan2010, angermann2014}. While the PZR is positive and linear as expected for p-Si, its magnitude is significantly larger than $\pi_{bulk}$ (see dashed, black line in Fig. \ref{preleased}(a)), with a maximum $\pi_{NW}\approx +3000\times 10^{-11}$ Pa$^{-1}$. Unlike the n-Si above, micro-Raman experiments reveal that this \textit{cannot} be explained by stress concentration alone (solid, gray line in Fig. \ref{preleased}(a)). Figure \ref{preleased}(b) shows the $\pi_{NW}/\pi_{bulk}$  ratio plotted against the experimentally measured value of $b$, each point corresponding to a single SiNW. $\pi_{NW}/\pi_{bulk}$ is now significantly larger than $b$ (solid, red line), despite the measured values of $b$ being close to the expected $b = 9$ (vertical, dashed line). Eq. \ref{effectivepi} cannot therefore describe the observed magnitude of the PZR.
 
In order to understand the excess PZR over and above that due to geometric stress concentration, an obvious starting point \cite{rowe2008, neuzil2010,jang2014,winkler2015} is the piezopinch effect arising from supposed electromechanical activity of the interface states \cite{rowe2008}. A stress-induced change in the total interface charge may arise if the activation energy of interface charge traps depends on stress \cite{hamada1994}, and will correspond to a change in the surface Fermi energy $E_{Fs}$. Via the requirement for charge neutrality this changes the width ($W$) of the surface depletion layer. In the limit of sufficiently small SiNW diameter $W \gg d$, the SiNW is strongly depleted of free charge carriers and $E_{Fs}$ imposes itself as the Fermi energy throughout the SiNW. Thus the stress will directly modify the free charge carrier density in the SiNW via a change in the interface trapped charge. This should be contrasted with the bulk PZR which is principally due to a change of the charge carrier mobility \cite{smith1954}. Although it is difficult to relate resistivity to a doping concentration (and hence a depletion layer width, $W$) in SiNWs \cite{schmidt2007}, the non-linear IV characteristic (inset, Fig. \ref{preleased}(b)) is a clear indication of the partial depletion necessary to observe a piezopinch effect. Partial depletion also ensures that the charge carriers may be treated classically (i.e. using Boltzmann statistics), and in this limit if the excess PZR is due entirely to a change in carrier concentration then \begin{equation} \label{shift} \Delta E_{Fs} = -k_BT \ln\left(1-\pi_{pinch}X\right) \end{equation} where $k_B$ is Boltzmann's constant and $T$ is the temperature. At most, the excess PZR over and above that described by geometric stress concentration is $\pi_{pinch} = \pi_{NW}/b - \pi_{bulk} \approx +260\times 10^{-11}$ Pa$^{-1}$. Using this value in Eq. \ref{shift} yields $\Delta E_{Fs} \approx 50$ $\mu$eV/MPa. Both the sign and magnitude of this change are in close agreement with measurements of charge trap activation energy shifts induced by mechanical stress in MOSFETs \cite{hamada1994, choi2008}. This offers tentative evidence for a link between the excess PZR and the stress-induced modulation of total interface charge.

In conclusion, it is unambiguously shown that the giant magnitude of the PZR in released, top-down SiNWs, which can be as large as the PZR initially reported in bottom-up SiNWs \cite{he2006}, is principally the result of geometric stress concentration. This observation opens the way to the integration of such elements into standard MEMS processes using lithography. Secondly, the central importance of surface chemistry in determining the sign or a smaller excess surface component of the PZR is demonstrated. Concentrated HF of the type typically used in MEMS processes is found to deplete Boron doped SiNWs, potentially changing the doping type and therefore yielding an apparently anomalous (negative) PZR. A poor quality oxide formed by rapid oxidation in concentrated HNO$_3$ of HF treated SiNWs yields p-type SiNWs whose PZR contains an excess surface component of magnitude similar to the bulk PZR. The magnitude is consistent with the known stress dependence of oxide trap activation energies, offering tentative evidence for a piezopinch-like PZR.

\acknowledgements{M.M. thanks the Labex NanoSaclay for financial support. This work was partially funded by the \textit{Agence Nationale de la Recherche} (ANR-10-NANO-0021) and the RENATECH network.}

\bibliographystyle{apsrev}

\bibliography{C:/Users/ar/Documents/Publications/References}
\end{document}